\documentclass[conference]{IEEEtran}
\IEEEoverridecommandlockouts
\usepackage{cite}
\usepackage{amsmath,amssymb,amsfonts,mathtools}
\usepackage{amsthm}
\usepackage{algorithmic}
\usepackage{graphicx}
\usepackage{textcomp}
\usepackage{xcolor}
\mathtoolsset{showonlyrefs}
\usepackage[left=0.65in,top=0.78in,right=0.65in,bottom=0.78in]{geometry}

\renewcommand{\baselinestretch}{1.03}
\def\BibTeX{{\rm B\kern-.05em{\sc i\kern-.025em b}\kern-.08em
    T\kern-.1667em\lower.7ex\hbox{E}\kern-.125emX}}
\begin{document}

\title{\huge Inter-Media Backscatter Communications with Magnetic Induction 	\vspace{-5pt}
}

\author{\IEEEauthorblockN{ Hongzhi Guo}
\IEEEauthorblockA{Engineering Department\\
	Norfolk State University\\
	hguo@nsu.edu 	}
\and
\IEEEauthorblockN{ Zhi Sun}
\IEEEauthorblockA{Electrical Engineering Department\\
	University at Buffalo, State University of New York\\
	zhisun@buffalo.edu 	}

}

\maketitle 
\vspace{-25pt}
\begin{abstract}

Wireless sensors in extreme environments, such as underground, concrete wall and the human body, can enable a large number of important applications. However, deploying wireless sensors on a large scale is a great challenge due to the environment and the high cost and large profile of wireless sensors. Backscatter communication can reduce the cost and size of wireless sensors by removing most of the typical wireless components. In this paper, we propose to leverage the RFID sensors for inter-media magnetic induction-based backscatter communications (MIBC). In this way, the complexity and cost of wireless sensors can be significantly reduced. The sensors leverage magnetic signals to backscatter information which demonstrate high penetration efficiency. We design a system with channel estimation, optimal signal transmission strategy, and an optimal receiver. The channel between the aboveground reader and underground sensors are modeled by using a stratified medium model. The bit-error-rate is evaluated with different configurations. The results suggest that MIBC can be utilized for most of the inter-media applications with low power consumption and high penetration efficiency.
\end{abstract}


\section{Introduction}

Ubiquitous and pervasive sensing in extreme environments finds a large number of civil and military applications. The underground sensors provide environmental information for  precision agriculture to improve farm yields \cite{salam2017based}. In-wall wireless sensors continuously monitor the strength of buildings. Intra-body wireless sensors can be leveraged to detect abnormalities at their early stages \cite{vasisht2018body}. However, the large profile and high cost of wireless sensors prevent us from deploying them in large scales. Moreover, replacing batteries in such inaccessible extreme environments is costly or even impossible. Therefore, an inter-media communication mechanism with low power consumption and small size is desirable.  

Backscatter communication (BC) is widely used in radio frequency identification (RFID) thanks to its low power consumption \cite{dobkin2012rf,boyer2014invited}. Recently, BC is also adopted to ambiently backscatter existing wireless signals, e.g., WiFi and DTV signals \cite{liu2013ambient,yang2015multi}, which is used in low-power internet of things (IoT). Sensors (e.g., temperature and image) can be integrated onto the RFID tag to provide sensing capability and the data is transmitted to the reader using BC \cite{liu2013ambient}. 

Although we can employ BC by deploying RFID sensors in an inaccessible medium (IM), the communication performances, e.g., bit error rate (BER) and power efficiency, are very limited since the IM is lossy, which absorbs wireless signal. Existing BC considers the UHF band, which has limited skin depth in underground, and thus the communication range is small. It requires high transmission power to compensate the high propagation loss. Moreover, the reflected signals by the air-soil boundary is much stronger than the signals scattered by RFID sensors. As a result, the reader cannot decode the scattered signals efficiently. In this paper, we adopt the magnetic induction (MI) communication, which is a low-power technology in extreme environments \cite{sun2010magnetic,guo2017multiple}. It leverages the HF (3-30MHz) or lower frequency bands to create a large skin depth to penetrate the inhomogeneous medium with negligible reflection and propagation loss. MI communications shares the same physical principles with the near field communications (NFC), which can be easily integrated to existing communications platforms. 

In this paper, we propose the magnetic induction backscatter communications (MIBC) for inter-media RFID sensors and, without loss of generality, we consider the IM is underground (UG) soil. The RFID sensors are employed to measure the moisture, conductivity, temperature and many other vital parameters. The RFID sensors demand energy for sensing, computation and data storage, which can be provided using wireless power transfer. The unmanned aerial vehicles (UAV) or mobile ground vehicles with readers can visit sensors regularly to collect sensed data. In this way, the sensors do not use active radios, which significantly reduces the energy consumption for wireless communications. The reader charges UG sensors using continuous waves (CWs), which can also be leveraged by the MIBC sensors to backscatter information.    

The contributions of this work are summarized as follows:
\begin{itemize}
	\item We bring the RFID sensor into inaccessible medium and use MIBC for data communications. To the best of our knowledge, this is the first paper investigating inter-media MIBC for lossy and inhomogeneous environment. 
	\item We design a comprehensive energy-efficient MIBC system, including channel estimation, signal modulation and detection, and optimum transmission strategy, which is a promising low-complexity solution for wireless sensing in IM environment.
	\item  We develop a tractable model using stratified medium model to prove the high penetration efficiency of MIBC across boundaries.  
\end{itemize}

The rest of this paper is organized as follows. In Section~II, we introduce the system model and the framework. After that, we provide the channel estimation strategy in Section III. The MIBC signal transmission and optimal detection strategies are designed in Section IV. Also, we develop an inter-media channel model by using the stratified medium model. The system performance is numerically simulated and evaluated in Section V. Finally, this paper is concluded in Section VI.  

The notations that are utilized in paper are summarized as follows: lower-case bold letters denote vectors and upper-case bold letters denote matrices. The subscript $t$ and $H$ stand for transpose and hermitian.The subscript $x_{a,r (i)}$ denote the real (imaginary) part of $x_a$. The complex number ${\tilde x}_{a}=x_{a,r}+jx_{a,i}$ is denoted with a tilde to distinguish it from real numbers. The superscript ${\tilde x}^{\ast}$ denotes the conjugate of a complex number ${\tilde x}$. 

\begin{figure}[t]
	\centering
	\includegraphics[width=0.42\textwidth]{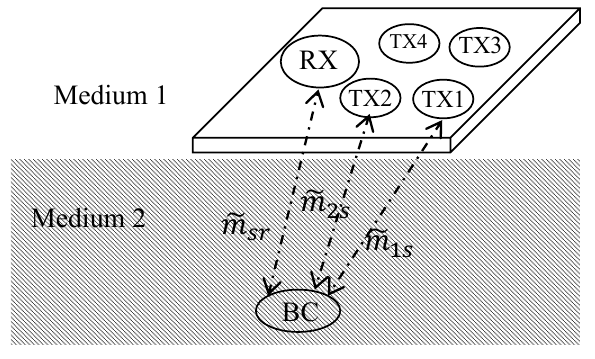}
	\vspace{-5pt}
	\caption{MIBC system with four transmit coils, one receive coil and one BC sensor. The channels between transmit coils 3, 4 and the UG sensors are not shown in the figure for ease of exposition and visualization.}
	\vspace{-10pt}
	\label{fig:sys}
\end{figure}
\section{System Model}

There are two key components in the proposed system, namely, the reader in the accessible medium, which can be a static reader or a mobile vehicle or a UAV, and the RFID sensors in IM. The reader has $N_t$ transmit coils and a receive coil. The multiple transmit coils are used to increase the diversity to overcome the orientation loss \cite{jadidian2014magnetic}. For example, when there is only one transmit coil, if the sensor coil orientation is perpendicular to the magnetic field generated by the transmit coil, there is no backscatter information since their coupling is zero. This problem can be addressed by using multiple transmit coils. If the sensor coil orientation is perpendicular to the magnetic field generated by one transmit coil, its coupling with other transmit coils cannot be zero due to the spatial diversity. We consider the reader is full-duplex using bistatic techniques to reduce the self-interference \cite{dobkin2012rf}. The reader is not a conventional RFID reader in the sense that it can not only read data from sensors but also wirelessly charge them using CWs. The operating frequency is 13.56MHz, which is widely used for RFID, NFC and wireless energy transfer. 

Each sensor has a single coil, an impedance modulator, a microprocessor and sensing units. Sensors can harvest energy from the reader, which is used for sensing, computing, data storage, and impedance modulation. Instead of generating high frequency signals, they transmit data by modulating the CWs sent by the reader. The modulation is conducted by changing coil impedance \cite{boyer2014invited,liu2013ambient}. In this way, the wireless component of the sensor is very simple and energy-efficient.

The operation of the system consists of two phases, namely, channel estimation and data transmission, as shown in Fig.~\ref{fig:estimation}. The reader can transmit signals with power $P_{h}$ and $P_{l}$, where $P_{h}>P_{l}$. The $P_{h}$ is used when there is no knowledge of the channel state information (CSI) between the sensor and reader, while the $P_{l}$ is used when CSI is available. The two different transmission strategies are introduced in detail in section III. We assume the sensor first harvest energy from the reader using magnetic induction, then it modulates signals to backscatter the CWs to the reader with information. In this paper, we only focus on the wireless communications. 

For the reader, we assume there are $N_t$ transmitters with the same impedance ${\tilde z}_t=r_t+j(\omega_c l_t-1/(\omega_c c_t))$, where $r_{t}$ is the resistance, $l_{t}$ is the self-inductance, $c_{t}$ is the capacitance to tune the circuit, and $\omega_c$ is the angular frequency. There is only one receiver with impedance ${\tilde z}_r=r_r+j(\omega_c l_r-1/(\omega_c c_r))$, where $r_r$, $l_r$, and $c_r$ are the resistance, self-inductance and capacitance, respectively. Each sensor has a coil with impedance ${\tilde z}_{s,p}=[r_s+j(\omega_c l_s-1/(\omega_c c_s))]e^{j\theta_p}$, where $p =1,2,\cdots,m$, $m$ is the symbol number, and $r_s$, $l_s$, and $c_s$ are the resistance, self-inductance and capacitance, respectively. The impedance ${\tilde z}_{sprd}=r_s+j(\omega_c l_s-1/(\omega_c c_s))$ is used for channel estimation and energy harvesting as a predefined symbol, which is known for both the reader and the sensor. 

Without loss of generality, we consider there is one active BC sensor in the IM. According to Kirchhoff's Law, the relation between voltages and currents can be written as ${\bf v}= {\bf Z} {\bf i}$, where ${\bf v}=[{\tilde v}_1, {\tilde v}_2,\cdots, {\tilde v}_{N_t}, \cdots, {\tilde v}_{N_t+2}]^t$ is the voltage, ${\bf i}=[{\tilde i}_1, {\tilde i}_2,\cdots, {\tilde i}_{N_t}, \cdots, {\tilde i}_{N_t+2}]^t$ is the current and ${\bf Z} \in {\mathcal C}^{(N_t+2)\times(N_t+2)}$, where the $i${th} diagonal element of ${\bf Z}$ is the impedance of the $i${th} coil and the off-diagonal element in the $i${th} row $p${th} column is $j\omega_c {\tilde m}_{ip}$, where ${\tilde m}_{ip}$ is the mutual inductance between the $i${th} coil and the $p${th} coil. The mutual inductance is a complex number due to the lossy underground environment. The transmit coils are numbered from $1$ to $N_t$, the BC sensor is $N_t+1$ and the receive coil is $N_t+2$. Since we consider the full-duplex reader, the ${\tilde m}_{k(N_t+2)}=0$ for $k=1,2,\cdots N_t$. Based on this model, we analyze the channel estimation and transmission strategies for MIBC.  
\section{Channel Estimation}
The reader relies on CSI to optimally transmit magnetic beams to reduce traveling loss and orientation loss. The CSI between the reader and the sensor is unknown and it changes with their positions. Moreover, compared with the far-field wireless communications, the effects of environment changes on MI channel are negligible. Hence, we do not need to perform channel estimation frequently as that in electromagnetic wave (EM)-based communications. In addition, there is no multipath fading in MI communications because signals are transmitted via induction; there is almost no propagation delay. As a result, we can safely consider that the channel has a single-tap. 

\begin{figure}[t]
	\centering
	\includegraphics[width=0.27\textwidth]{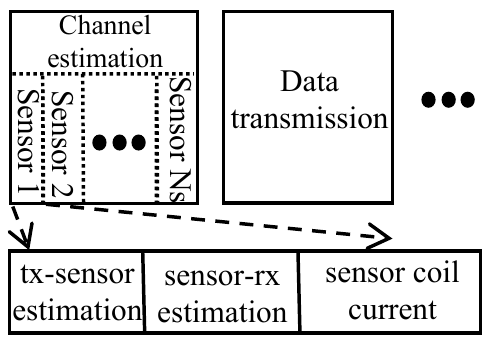}
	\vspace{-5pt}
	\caption{Channel estimation protocol for MIBC.}
	\vspace{-10pt}
	\label{fig:estimation}
\end{figure}

MI channel estimation has been investigated in \cite{kisseleff2014transmitter,jadidian2014magnetic,moghadam2017node}. Different from existing works, the BC has a dyadic backscatter channel (DBC)\cite{boyer2014invited}, which is challenging to estimate since UG sensors cannot process the received signals without active radios components, such as oscillators. For a DBC, the downlink is integrated with the uplink, which makes the channel estimation difficult in EM-based communications \cite{yang2015multi}. In this paper, we propose a channel estimation protocol, which is shown in Fig.~\ref{fig:estimation}. Thanks to the stable channel, the results of channel estimation can be utilize for data transmissions within a long period provided that the reader does not change its position and there is no abrupt environmental change. 

Although we consider one sensor here, the channel estimation for multiple sensors can be performed in a random access fashion. Since the sensors can only blindly send their measurements and there is no feedback from the reader, collisions cannot be detected or avoided by sensors. The reader can estimate the number of sensors and keep sending CWs until all the required information are received correctly \cite{wang2012efficient}. The UG sensors are synchronized; they compete to reserve a time lot. If a sensor wins, all other sensors keep silent in that time slot. We introduce the estimation procedure in the following.

The channel estimation for a single sensor consists of three phases. First, in the transmitter-sensor (${\tilde m}_{ts}$) estimation, we turn down the receiver and let the transmitter send CWs to the sensor with the high transmission power $P_h$. Each coil of the transmitter is allocated with the same amount of transmission power, i.e., $P_h/N_t$. The sensor measures the current in its coil. Meanwhile, each transmit coil measures its voltage and current, which are saved for future processing. Assume the transmitted CW in the transmit coil $k$ is ${ i}_k(t)=\Re\{{\tilde i} _ke^{j\omega_c t}\}$, which results in a voltage across the sensor coil, 
\begin{align}
\label{equ:sensorest}
{v}_{s}(t)&=\Re\{{\tilde v}_se^{j\omega_c t}\}=\Re\{(-\sum_{k=1}^{N_t}{\tilde i}_kj\omega_c {\tilde m}_{ks} +{\tilde n}_s)e^{j\omega_c t}\}\nonumber\\
&\approx\Re\{(-\sum_{k=1}^{N_t}{\tilde i}_kj\omega_c {\tilde m}_{ks} )e^{j\omega_c t}\},
\end{align} 
where $ {\tilde m}_{ks} $ is the channel coefficient (mutual inductance) between the transmit coil $k$ and the sensor coil,  and ${\tilde n}_s$ is the noise. Note that, since the communication is based on backscatter, ${\tilde n}_s$ is much smaller than the voltage induced by transmitted CWs, otherwise the receiver can barely detect backscattered signals. In the following we only keep the phasor to simplify the notations. 

Second, in the predefined sensor-receiver (${\tilde m}_{sr}$) estimation phase, the receiver is turned on and the sensor transmits predefined data, i.e., use predefined impedance. Then, the voltage across the receiver's coil is
\begin{align}
\label{equ:vrx}
{\tilde v}_{r}=\frac{-\sum_{k=1}^{N_t}{\tilde i}_k\omega_c^2 {\tilde m}_{ks}{\tilde m}_{sr}}{{\tilde z}_{sprd}}+{ \tilde n}_{r};
\end{align} 
where ${\tilde z}_{sprd}$ is the predefined impedance that the sensor uses in this time frame and ${\tilde n}_r\in {\mathcal {CN}}(0,\sigma_r^2)$ is the noise in the receiver. Note that, in this process, the sensor only adjusts its impedance to ${\tilde z}_{sprd}$ and the reader measures the voltage ${\tilde v}_{r}$ to estimate ${\tilde m}_{ks}$ and ${\tilde m}_{sr}$. 

Third, in the sensor current transmission phase, the sensor sends the measured ${\tilde v}_s$ in the first phase to the receiver using differential phase shift keying (DPSK) modulation, which will be introduced in next section. The current and voltage are linearly related ${\tilde i}_s={\tilde v}_s/{\tilde z}_{s}$, upon which it can estimate the CSI (or mutual inductance) of the channels. The procedures are given as follows. 

First, for the $N_t$ transmit coils we have
\begin{align}
\label{equ:estimate1}
&{\bf  v}_{N_t}={\bf  Z}_{N_t}
{\bf  i}_{N_t}+j \omega_c {\tilde i}_{s}{\bf  m}_{ts},
\end{align}
where ${\bf  v}_{N_t}$ and ${\bf  i}_{N_t}$ are the first $N_t$ elements of ${\bf v}$ and ${\bf i}$ given in the system model, respectively; ${\bf Z}_{N_t}$ is the first $N_t$ column and first $N_t$ row of ${\bf Z}$; and $ {\bf  m}_{ts}$ is the unknown mutual inductance between the transmit coils and the sensor coil.
Therefore, the mutual inductance can be estimated by 
\begin{align}
\label{equ:es_hts}
{\hat {\bf  m}}_{ts}=\frac{1}{j\omega_c {\tilde i}_s}\left({\bf v}_{N_t}-{\bf Z}_{N_t}
{\bf i}_{N_t}\right).
\end{align}
 Using \eqref{equ:vrx}, the CSI of the channel between the sensor coil and the receive coil can be estimated by 
\begin{align}
\label{equ:es_hsr}
{\hat {\tilde m}}_{sr} = \frac{-{\tilde v}_{r}{\tilde z}_{sprd}}{\sum_{k=1}^{N_t}\omega_c^2{\tilde i}_k{\hat {\tilde m}_{ts}}},
\end{align}
The estimation error of ${\hat {\tilde m}}_{sr} $ is
\begin{align}
\label{equ:erres_hsr}
{\tilde e_{sr}}={\tilde m}_{sr}-{\hat {\tilde m}}_{sr}= {\frac{{\tilde n}_r {\tilde z}_{sprd}}{\sum_{k=1}^{N_t}\omega_c^2{\tilde i}_k{\hat {\tilde m}}_{ts}}}={\tilde \beta}({\tilde m}_{ts}){\tilde n}_r,
\end{align}
where ${\tilde \beta}({\tilde m}_{ts})= {{\tilde z}_{sprd}}/{\sum_{k=1}^{N_t}(\omega_c^2{\tilde i}_k{\hat {\tilde m}}_{ts})}$, which is a constant once the transmission strategy is determined. As a result, ${\tilde e_{sr}}$ is also a complex normal distribution, i.e., ${\mathcal {CN}}(0, |{\tilde \beta}({\tilde m}_{ks})|^2\sigma_r^2)$. In view of \eqref{equ:erres_hsr}, to reduce the estimation error we can increase the transmission power (increase ${\tilde i}_k$) or reduce the impedance of the sensor coil ${\tilde z}_{sprd}$. Recall that we use $P_h$ in channel estimation; the reason is that we can reduce the estimation errors in this way. Up to this point, all the CSI have been found in \eqref{equ:es_hts} and \eqref{equ:es_hsr}, upon which we can design the optimal signal detection and transmission strategy.
\section{Optimal Transceiver Design for MIBC}
In this section, we introduce the signal detection approach and wireless transmission strategy. Also, we show the advantages of MIBC compared with existing solutions, especially the electromagnetic wave-based solutions.  

\subsection{Receiver Detection Design}
To increase the reliability and power-efficiency, here we consider the DPSK modulation with $m$ symbols $\{{\tilde y}_{s,1}, {\tilde y}_{s,2},\cdots, {\tilde y}_{s,m}\}$, where ${\tilde y}_{s,p}=1/{\tilde z}_{s,p}$ is the admittance of the $p${th} coil that is used to modulate the CW. The CW is penetrated into the inaccessible medium. Since we use magnetic induction, the reflection from the boundary is negligible, which will be proved in Section IV-C. Next, we analyze the modulation and demodulation in sequel.  

First, in the sensor coil we have the signal
\begin{align}
\label{equ:immodulation}
{ \tilde v}_s=
-j\omega_c\sum_{k=1}^{N_t} {\tilde i}_k {\tilde m}_{ks}-j\omega_c {\tilde i}_r {\tilde m}_{sr}.
\end{align}
Here the noise is neglected due to the large induced voltage in the sensor coil. Also, \eqref{equ:immodulation} is different from \eqref{equ:sensorest}, since in the channel estimation all sensors have their own time slots and thus there is no interference. In \eqref{equ:immodulation}, the first term on the right-hand-side (RHS) is the voltage induced by the transmit coils, while the second term is the voltages induced by the receive coil. The first term is much larger than the other one because $i_r$ is induced by the sensor coil current rather than being actively generated by the receive coil. Thus, we can approximately obtain ${ \tilde v}_s\approx
-j\omega_c\sum_{k=1}^{N_t} {\tilde i}_k {\tilde m}_{ks}$. Based on it we can obtain the induced voltage in the receive coil when the transmit symbol is ${\tilde y}_{s,p}$,
\begin{align}
\label{equ:immodulation2}
{\tilde v}_{r,p}={\tilde g}{\tilde y}_{s,p}+{\tilde n}_r,
\end{align}
where ${\tilde g}=-\sum_{k=1}^{N_t}\omega_c^2 {\tilde i}_k {\tilde m}_{ks} {\tilde m}_{sr}$.

To recover the transmitted symbol ${\tilde y}_{s,p}$, we rewrite the received signal model as 
\begin{align}
\label{equ:demo1}
{\tilde v}_{r,p}=|{\tilde g}|e^{j\phi_g}{\tilde y}_{s,p}+{\tilde n}_r,
\end{align}
where $|{\tilde g}|$ and $\phi_g$ are the magnitude and phase of ${\tilde g}$, respectively. Referring to \eqref{equ:immodulation2}, $\phi_g$ is a function of ${\tilde m}_{sr}$, which cannot be estimated without any error. Therefore, we adopt a noncoherent detection design by considering $\phi_g$ as a random variable with uniform distribution in $[0, 2\pi]$. 
 Assume the transmitted signal is $\{{\tilde y}_{s}\}=[t_1,t_2]=[{\tilde y}_{s,p},{\tilde y}_{s,p}e^{j\phi_m}]$ and the received signal is
\begin{align}
\{{ {\tilde v}}_{r,p}\}=[r_1,r_2]=|{\tilde g}|e^{j\phi_g}[{\tilde y}_{s,p},{\tilde y}_{s,p}e^{j\phi_m}]+[{\tilde {n}}_{r1},{\tilde {n}}_{r2}]
\end{align}
Then, the optimal detector can be written as
\begin{align}
\label{equ:detector1}
{\hat p} = \underset{1\leq p \leq m}{\arg \max}~\frac{P_p}{2\pi}\int_{0}^{2\pi} {\mathbb P}(\{{\acute {\tilde v}}_{r,p}\}-|{\tilde g}|e^{j\phi_g}\{{\tilde y}_{s,p}\}) d\phi_g,
\end{align}
where $P_p$ is the probability that the symbol $p$ was transmitted. The solution to \eqref{equ:detector1} is when the magnitude of the product between the symbol and the received signal is maximized \cite{proakis2008digital}, which is
\begin{align}
\label{equ:demo3}
{\hat p} = \underset{1\leq p \leq m}{\arg \max}~|\{{\acute {\tilde v}}_{r,p}\}\cdot\{{\tilde y}_{s,p}\}|=\underset{1\leq p \leq m}{\arg \min}|\angle r_2-\angle r_1 -\phi_m|.
\end{align}
The detector has a simple structure for implementation.  
The signal-to-noise ratio (SNR) which is 
$
\label{equ:snr}
{\mathcal T}=\frac{|{\tilde g}{\tilde y}_{s,p}|^2}{\sigma_r^2}.
$
The data rate $R$ of the proposed MIBC is determined by the joint bandwidth $W$ of coil antennas and MI channels \cite{guo2017multiple}, and the relation is $R=W\log_2 m$ for low BER. As a result, for DBPSK $R=W$, while for DQPSK $R=2W$.  
\subsection{Wireless Transmission Strategy}
To successfully receive the transmitted information, the transmit currents have to be controlled in order to obtain the SNR to achieve the required BER. Here, we try to maximize the received SNR given the limited transmit power
\begin{align}
\label{equ:received_power}&{\max}~~{\mathcal T} \\
\label{equ:constrains}&{\text {subject to}}  \sum_{k=1}^{N_t}P_{tk} \leq P_{l},
\end{align}
where $\tau_1$ is the required SNR and $P_{tk}$ is the transmit power of the $k$th transmit coil. 
To find the optimal transmission solution, we change the variables to the currents of transmit coils. The SNR can be written in matrix form as 
\begin{align}
{\mathcal T}=\frac{{\bf { i}}_t^H {\bf C}_1 {\bf { i} }_t}{\sigma_r^2},
\end{align}
where ${\bf C}_1={\bf { m}}_{ts}^{H}{\bf { m}}_{ts}|\omega_c^2 {\hat {\tilde m}}_{sr} {\tilde y}_{s,p}|^2$. Although ${\bf C}_1$ consists of the symbol ${\tilde y}_{s,p}$, different symbols do not affect the optimal transmission strategy because they share the same magnitude. 
The transmit power of a coil is $P_{tk}={|{\tilde i}_{tk}|^2r_t}/{2}$ and the constraint in \eqref{equ:constrains} can be expressed as
\begin{align}
\label{equ:opt3}
{\bf {i}}_t^H {\bf Z}_{N_t} {\bf { i} }_t\leq P_t
\end{align}
where ${\bf Z}_{N_t}$ is given in \eqref{equ:estimate1}. 
Since the original problem is nonconvex, a solution cannot be efficiently found. We reorganize the problem by considering the the SNR as a constrain and minimize the transmit power, i.e.,
\begin{align}
&\underset{{\bf i}_t}{\min} ~~{\bf {i}}_t^H {\bf Z}_{N_t} {\bf { i} }_t \nonumber\\
& {\text {subject to}}~~{\bf { i}}_t^H {\bf C}_1 {\bf { i} }_t\geq {\mathcal T}\sigma_r^2.
\end{align}
Then, the above problem becomes a nonconvex quadratically constrained quadratic programming. Since the objective function is convex (${\bf Z}_{N_t} \succeq0$), it can be solved efficiently using semidefinite relaxation \cite{luo2010semidefinite}. The optimal ${\bf { i}}_t$ can satisfy the SNR constrain and minimize the transmitted power. From the perspective of implementation, the current can be generated using a feedback loop, i.e., by measuring the magnitude and phase of the current in a transmit coil and compare it with the expected value. 
\subsection{Effects of Inhomogeneous and Lossy Media}  
 Up to this point, we have developed the communication system by assuming the CSI can be estimated. In this subsection, we investigate the effects of the inhomogeneous and lossy UG environment and develop a mutual inductance model by considering the air-soil interface as well as the soil conductivity.  Moreover, the advantage of MIBC over EM-based communications in UG environment is discussed. 

\begin{figure}[t]
 	\centering
 	\includegraphics[width=0.25\textwidth]{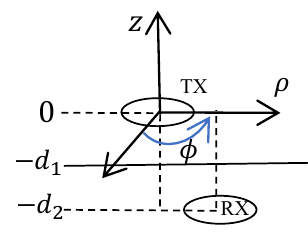}
 	\vspace{-5pt}
 	\caption{Stratified medium model in Cylindrical Coordinates. The upper layer (above -$d_1$) is the air (accessible medium) and the lower layer (below -$d_1$)  is the UG soil (inaccessible medium).}
 	\vspace{-10pt}
 	\label{fig:stratified}
 \end{figure}

 The EM solution leverages UHF band (300MHz to 3GHz) signals \cite{salam2017based} to penetrate through the soil medium. Field experiments have demonstrated the feasibility of this solution. However, the wireless signals experience significant power loss when propagate through soil; this problem becomes severe in lossy soil medium since a large amount of power can be absorbed by the soil. To provide a rigorous comparison, we develop a comprehensive model for the propagation loss of MI and EM by using a stratified medium model, as shown in Fig.~\ref{fig:stratified}. 
 
 \subsubsection{Propagation Analysis}
 A vertically-orientated coil with time varying current $i_t$ generates magnetic field ${\bf h}=(h_{\rho},h_{\phi},h_z)$ in space, which is modeled in Cylindrical Coordinates. The z-direction magnetic field in air is \cite{chew1995waves}
 \begin{align}
h_{ag,z}=-\frac{ji_t\pi a^2 N_{c}}{8\pi}\int\limits_{-\infty}^{\infty}dk_{\rho}\frac{k_{\rho}^3}{k_{1z}}H_0^{(1)}(k_{\rho}\rho)\left(e^{jk_{1z}|z|}\right.\nonumber\\
 \left.+{\mathcal R}_{12}e^{jk_{1z}z+2jk_{1z}d_1}\right),
 \end{align} 
 where $a$ is the coil radius, $N_{c}$ is the coil number of turns, $k_{1z}=\sqrt{k_1^2-k_{\rho}^2}$, $k_1=\omega_c\sqrt{\mu_1\epsilon_1}$, $\mu_1$ is the permeability of air, $\epsilon_1$ is the permittivity of air, $\rho$ is the horizontal distance in cylindrical coordinates, $H_0^{(1)}(x)$ is the first kind of Hankel function with order 0, and ${\mathcal R}_{12}$ is the reflection coefficient of the air-soil interface. The UG magnetic field is 
  \begin{align}
  \label{equ:hugz}
 h_{ug,z}=-\frac{ji_t\pi a^2 N_{c}}{8\pi}\int\limits_{-\infty}^{\infty}dk_{\rho}\frac{k_{\rho}^3}{k_{2z}}H_0^{(1)}(k_{\rho}\rho){\mathcal T}_{12}e^{-jk_{2z}z},
 \end{align} 
 where $k_{2z}=\sqrt{k_2^2-k_{\rho}^2}$, $k_2=\omega_c\sqrt{\mu_2\epsilon_2}$, and $\mu_2$ and $\epsilon_2$ are the permeability and permittivity of the soil, respectively. Note that, for lossy soil we consider the complex permittivity, i.e., $\epsilon_{2}=\epsilon_{soil}+j\sigma_2/\omega_c$ with soil conductivity $\sigma_2$ and permittivity $\epsilon_{soil}$. The transmission coefficient for a vertically orientated coil is
 \begin{align}
 {\mathcal T}_{12}=\frac{2\mu_2 k_{1z}}{\mu_2k_{1z}+\mu_1k_{2z}}e^{jd_1(k_{1z}-k_{2z})}. 
 \end{align}
 By using \eqref{equ:hugz}, we can obtain the transverse components
 \begin{align}
 (h_{ug,\rho},h_{ug,\phi})=\int\limits_{-\infty}^{\infty}\frac{1}{k_{\rho}^2}\left(\frac{\partial}{\partial \rho}\frac{\partial  h_{ug,z}}{\partial z},\frac{1}{\rho}\frac{\partial}{\partial \phi}\frac{\partial  h_{ug,z}}{\partial z}\right)dk_{\rho}\nonumber \\ =\frac{i_t\pi a^2 N_{c}}{8\pi}\int\limits_{-\infty}^{\infty}\left(k_{\rho}^2H_1^{(1)}(k_{\rho}\rho),0\right) {\mathcal T}_{12}e^{-jk_{2z}z}dk_{\rho}.
 \end{align} 
 Once we have the intensity of the magnetic field ${\bf h}_{ug}= (h_{ug,\rho},h_{ug,\phi},h_{ug,z})$ in underground, the mutual inductance between a coil in air and a coil in soil is 
 \begin{align}
 \label{equ:mutual}
 m_{ag,ug}=\frac{\mu_2\pi a^2N_c {\bf h}_{ug} {\bf n}_c}{i_t},
 \end{align}
 where ${\bf n}_c$ is a unit vector denoting the orientation of a coil. The propagation efficiency is defined as \cite{lin2015distributed}
 \begin{align}
 \eta_{mi}=\frac{p_{rec}}{p_{tra}}=\frac{\omega_c^2  |m_{ag,ug}|^2}{2(\omega_c^2  |m_{ag,ug}|^2+2r_s^2)}.
 \end{align}   
 Note that, since we focus on the propagation loss, we implicitly assume the coils are perfectly matched to avoid antenna reflection losses. 
 
 Next, we study the propagation loss of EM solutions. Since in UHF the wavelength is much smaller compared with the distance between the devices in air and soil, here we employ the wave propagation model instead of full EM fields analysis.
 To remove the effects of incident angle, we assume the coil in the soil is right under the the coil in the air, i.e., there is no horizontal deviation. Then, the propagation efficiency using the received power divided by the transmitted power can be written as
 \begin{align}
  \eta_{em}=\left(\frac{\lambda_1}{4\pi d_1}\right)^2\left(\frac{\lambda_2}{4\pi d_2}\right)^2G_1G_2{\mathcal T}_{em}e^{-2k_{2,i}d_2},
 \end{align} 
 where $\lambda_i$ and $G_i$ are the wavelength and antenna gain in the $i${th} medium, respectively. ${\mathcal T}_{em}=2\eta_2/(\eta_2+\eta_1)$, where $\eta_i=\sqrt{\epsilon_i/\mu_i}$. By using the developed model, we compare the performance of MIBC and existing solutions in next section. 
\subsubsection{Reflection Analysis}
For backscatter communications in inhomogeneous medium, the reflection can be considered as interference with reasonable strength since the reflection from the boundary travels a shorter distance than the backscattered signals. It is a great challenge for UHF signals since the wavelength is small and the reflection is strong on the boundary. As a result, the receiver receives a large reflected signal, which can be much stronger than the backscattered signals. This problem can be solved by using MIBC, because the magnetic field has strong penetration efficiency and thus negligible reflections. The reflection coefficient for a small magnetic coil, which is close the boundary, can be written as
\begin{align}
{\mathcal R}_{12}=\frac{k_{1z}-k_{2z}}{k_{1z}+k_{2z}}e^{2jk_{1z}d_1}.
\end{align}  
Since the evanescent wave is dominant the near field, $k_{\rho}$ is much larger than $k_1$ and $k_2$. As a result, $k_{1z}\approx k_{2z}$ and, hence, the reflection can be neglected. 
\section{Performance Evaluation}
In this section, we numerically simulate the performance of the proposed system. In \eqref{equ:mutual}, we provide an equation for the mutual inductance between two coils in different media. For the coils of the reader, they are very close to each other and the coils cannot be regarded as a dipole point. Here, we adopt an exact solution and the mutual inductance between the transmit coil $p$ and the transmit coil $q$ can be written as \cite{conway2007inductance}
\begin{align}
m_{p,q}=\mu_1\pi a_pa_qN_pN_q\int\limits_{0}^{\infty} J_0(sd_{pq})J_1(sa_p)J_1(sa_q)ds,
\end{align} 
where $a$ is the coil radius, $N$ is the coil number of turns, $d_{pq}$ is the distance between the two coils, and $J_n(x)$ is the Bessel function of the first kind with order $n$. To numerically evaluate the above equation is challenging since it consists of oscillating functions. Here, we adopt the Gauss Quadrature to find the solutions, which has the same accuracy as the results in \cite{conway2007inductance}. 

The parameters are given as follows. The reader has a 0.15m$\times$0.1m board to accommodate the transmit and receive coils. The radii and number of turns of the four transmit coils and the receive coil are 0.02m and 5, respectively. By using copper wires with radius 0.145cm, the coil resistance, self-inductance and the tuning capacitance are 0.13$\Omega$, 1.69$\mu$H and 81.51pF, respectively. The sensor coils has 2 turns with radius 0.02m to reduce its size and weight, and the coil resistance, self-inductance, and the tuning capacitance are 0.013$\Omega$, 0.1$\mu$H and 81.51pF, respectively. All the coils are vertically orientated (z-axis). The real relative permittivity, relative permeability and conductivity of soil are 5, 1 and 0.01S/m, respectively. The frequency is 13.56MHz. The noise level is considered as -80dBmV. The reader is 0.05m above the air-soil interface, i.e., $d_1=0.05$m. We consider the modulation schedule is DQPSK with four symbols. 

	\begin{figure}[t]
	\centering
	\includegraphics[width={0.4\textwidth}]{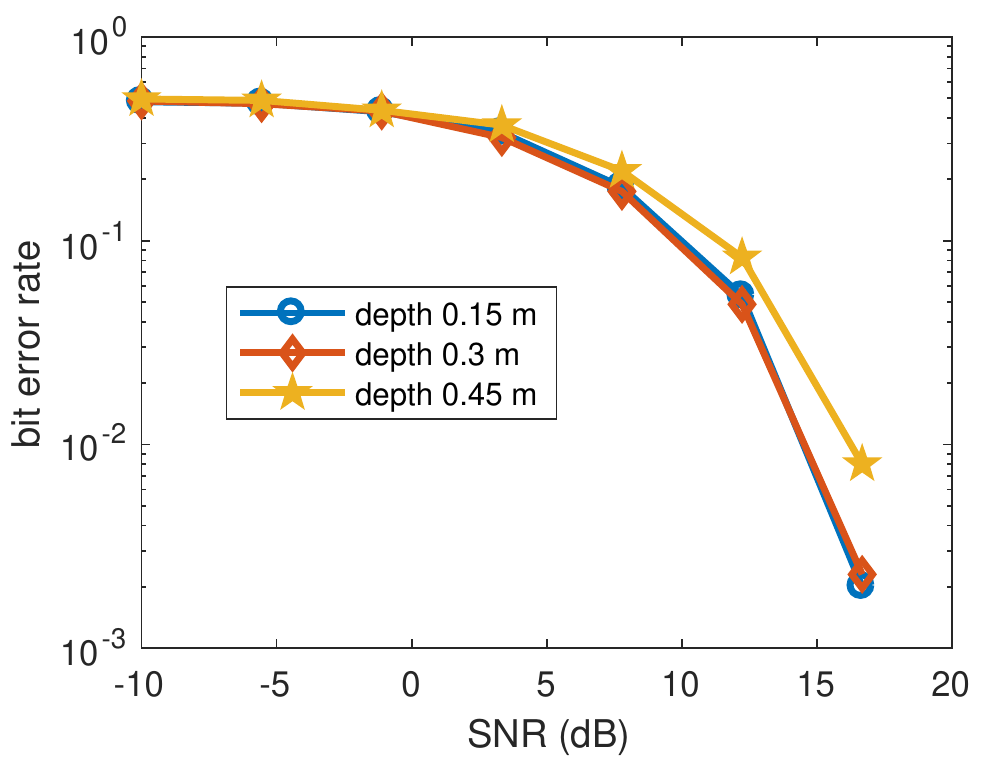}
	\caption{Effect of sensor depth. The horizontal distance between the reader and the sensor is 0.5m.  }
	\label{fig:ber}
	\vspace{-5pt}
\end{figure}

\begin{figure}[t]
	\centering
	\includegraphics[width={0.37\textwidth}]{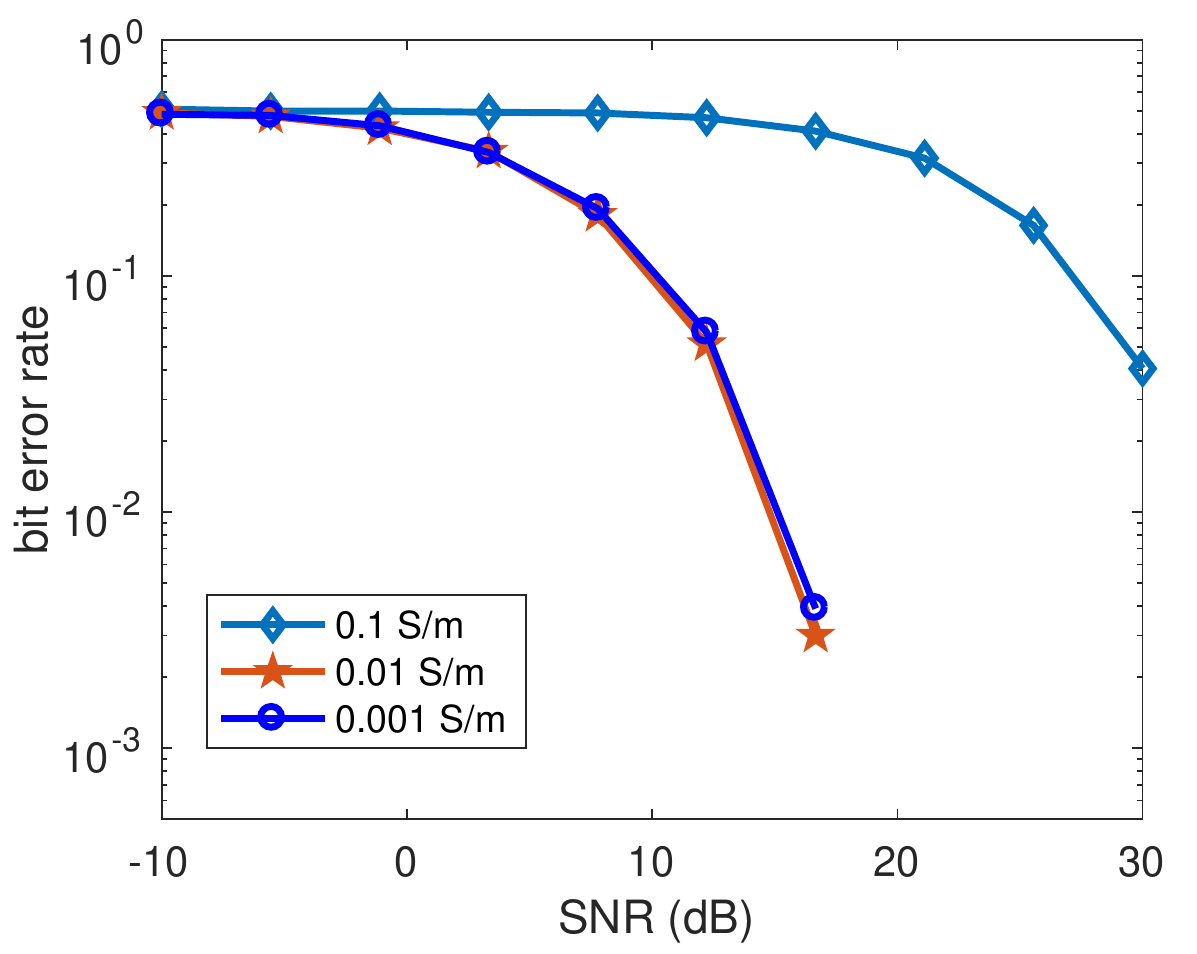}
	\caption{Effect of soil conductivity.  }
	\label{fig:pr}
	\vspace{-5pt}
\end{figure}

\begin{figure}[t]
	\centering
		\includegraphics[width={0.43\textwidth}]{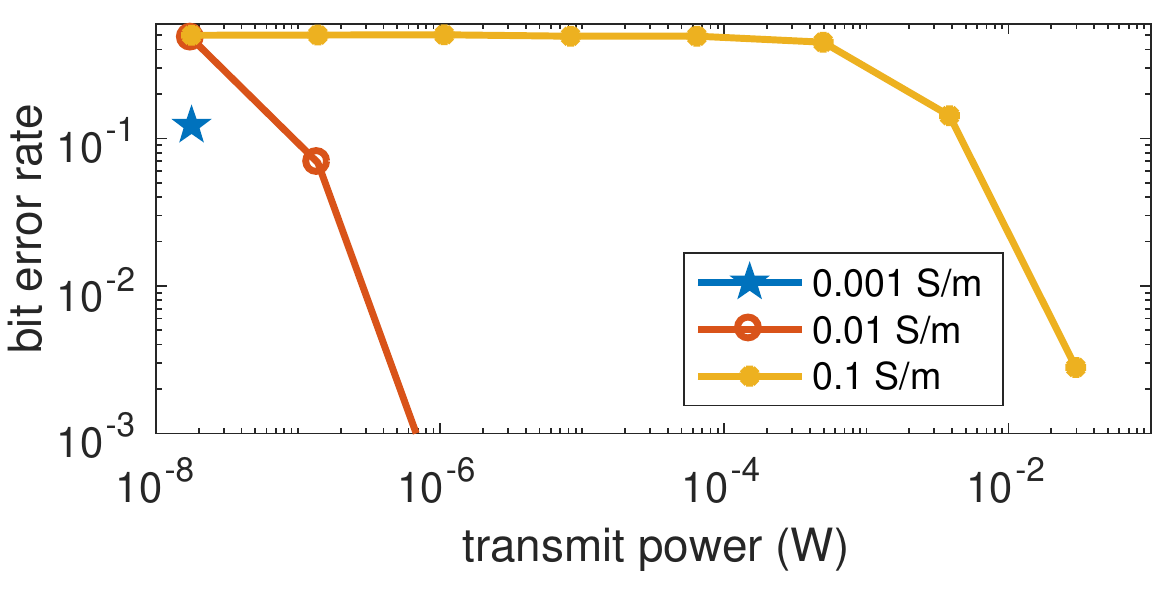}
		\caption{Effect of the transmit power for channel estimation with various underground soil conductivity. }
		\label{fig:estimation_power}
				\vspace{-5pt}
	\end{figure}

	\begin{figure}[t]
			\centering
		\includegraphics[width={0.35\textwidth}]{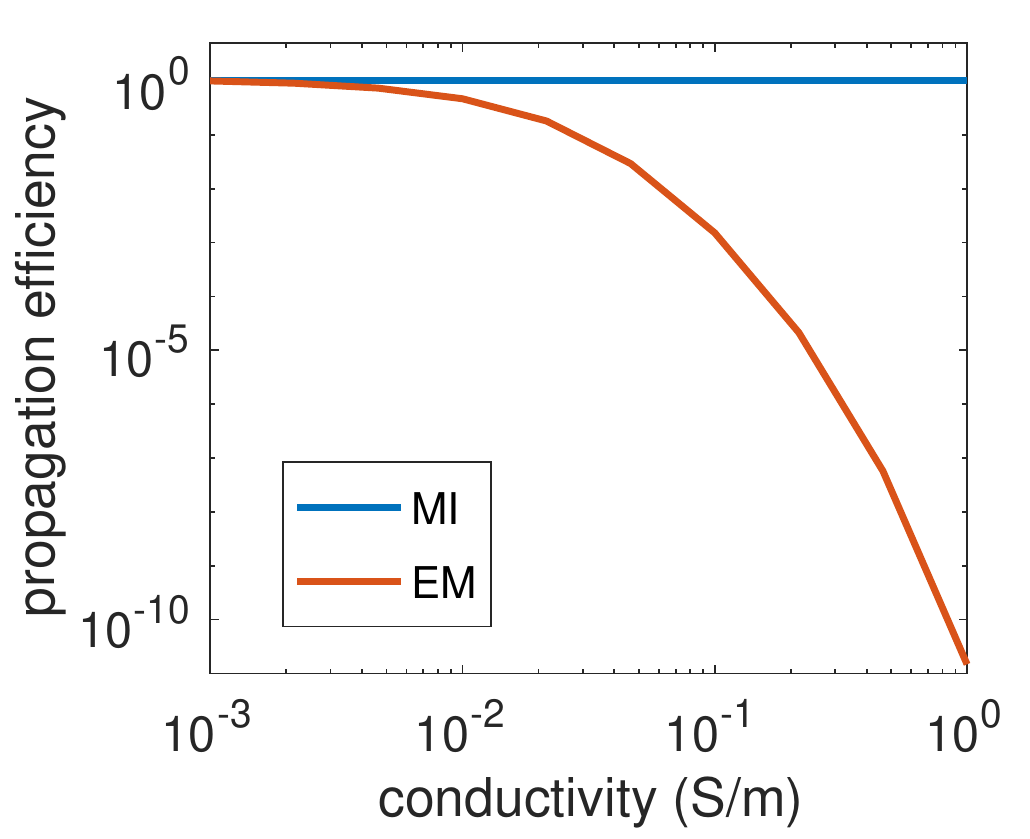}
		\caption{Comparison of propagation efficiency through lossy medium for MI and EM communications. }
		\label{fig:loss}
		\vspace{-5pt}
\end{figure}
First, the effect of sensor depth is shown in Fig.~\ref{fig:ber}. The horizontal distance between the reader and the sensor is 0.5m. We change the vertical depth of the sensor from 0.15m to 0.45m, which are typical depth for most of the underground agriculture applications. The SNR is increased from -10dB to 20dB. We run the simulation for 10,000 times to calculate the BER. As shown in the figure, when the SNR is higher than 16dB the BER becomes zero for all the considered depth. When the depth is 0.45m, the performance is a little worse, but very close the other smaller depth, which shows the robustness of the MIBC.

The effect of soil conductivity is shown in Fig.~\ref{fig:pr}. The horizontal distance does not change and the vertical depth is 0.3m. As shown in the figure, only very high conductivity, i.e., 0.1S/m, can generate strong effects. The MIBC cannot work in such highly lossy medium due to the significant signal attenuation. The sensor does not have active input and the path loss is large. But when the conductivity is small, which is true for most underground soil, the performance is reliable. 


Channel estimation error can affect the optimal transmission strategy and the signal detection. In Fig.~\ref{fig:estimation_power}, we change the transmission power for channel estimation and show the BER of the received signals. As shown in the figure, when the power is low, the channel estimation error is high and all information we use for signal detection and optimal transmission are not accurate. As a result, the BER is large. Also, the low conductivity medium requires smaller transmit power. As conductivity increases, higher transmit power is required to compensate the absorption of the lossy medium. In Fig.~\ref{fig:loss}, we gradually increase the conductivity of the soil medium and show the propagation efficiency derived in Section IV-C. Here we normalized the propagation efficiency by dividing it by the efficiency when soil conductivity is 1mS/m to ease the exposition and visualization. The results suggest that when the environment changes, MI communication is robust, while the EM-based communication suffers from a dramatic loss in efficiency. As a result, in dynamic extreme environments, MI is a better solution. 

 In addition, compared with point-to-point active MI communications \cite{kisseleff2014transmitter,guo2017multiple}, the MIBC does not actively generate wireless signals and thus its power consumption is low. Different from wireless sensors, the main power consumption of BC sensor are the sensing, data storage, and computation, which occupies less than 20\% of the overall power consumption in wireless sensors \cite{akyildiz2010wireless}. Therefore, the power efficiency can be increased dramatically.

\section{Conclusion}
A magnetic induction-based backscatter communication (MIBC) system for inter-media wireless sensing is designed and analyzed in this paper. The low-cost and low-profile RFID sensor is buried in inaccessible medium to monitor various parameters and backscatter the sensed data. Thanks to the high penetration efficiency of magnetic induction communications, the reflected signals on the boundary can be neglected. We design a complete communication system, including channel estimation, modulation, demodulation, optimal transmission strategy, and developed a comprehensive underground MI channel model for inhomogeneous underground medium to characterize the signal propagation. The results show that the MIBC consumes low power and it is feasible to work in most inaccessible medium without high conductivity. The multiple transmit coils provide flexibility and reliability to the system, upon which we can optimally control the backscatter communications. The developed system can find many other applications in extreme environment, such as in-wall structure monitoring and intra-body sensing.
\bibliographystyle{IEEEtran}
\bibliography{guo}
\end{document}